\documentclass[12pt]{article}
\usepackage{epsfig}
\usepackage{a4}

\begin{document}  
\renewcommand{\thefootnote}{\fnsymbol{footnote}}
\setcounter{page}{0}
\begin{titlepage}   
\vspace*{-2.0cm}  
\begin{flushright}
\end{flushright}
\vspace*{0.1cm}
\begin{center}
{\Large \bf SNO+: predictions from standard solar models and resonant spin flavour 
precession} 
\vspace{0.6cm}

\vspace{0.4cm}

{\large 
Marco Picariello$^{a\,b}$ \footnote{E-mail: Marco.Picariello@le.infn.it},
Jo\~{a}o Pulido$^a$\footnote{E-mail: pulido@cftp.ist.utl.pt}\\
\vspace{0.15cm}
{  {\small \sl
$^a$ Centro de F\'{\i}sica Te\'{o}rica das Part\'{\i}culas (CFTP) \\
 Departamento de F\'\i sica, Instituto Superior T\'ecnico \\
Av. Rovisco Pais, P-1049-001 Lisboa, Portugal\\
\vspace{0.25cm}
$^b$ I.N.F.N. - Lecce, and Dipartimento di Fisica\\
Universit\`a di Lecce
\\ Via Arnesano, ex Collegio Fiorini, I-73100 Lecce, Italia\\
}}}
\vspace{0.25cm}
{\large 
S. Andringa, N.F. Barros, J. Maneira\\
\vspace{0.15cm}
{  {\small \sl Laborat\'orio de Instrumenta\c{c}\~{a}o e F\'\i sica Experimental de Part\'\i culas,\\
 Av. Elias Garcia, 14, 1$^\circ$, 1000-149 Lisboa, Portugal}
}}
\vspace{0.25cm}
\end{center}
\vglue 0.6truecm
\begin{abstract}
Time variability of the solar neutrino flux especially in the low and intermediate 
energy sector remains an open question and, if it exists, it is likely to be
originated from the magnetic moment transition from active to light sterile 
neutrinos at times of intense solar activity and magnetic field.
We examine the prospects for the SNO+ experiment to address this important issue and 
to distinguish between the two classes of solar models which are currently identified 
as corresponding to a high (SSM~I) and a low (SSM~II) heavy element abundance. 
We also evaluate the predictions from these two models for the 
Chlorine experiment event rate in the standard LMA and LMA+Resonant Spin Flavour 
Precession (RSFP) scenarios.
It is found that after three years of SNO+ data taking, the pep flux measurement will
be able to discriminate between the standard LMA and LMA+RSFP scenarios, independently 
of which is the correct solar model. If the LMA rate is measured, RSFP with $B_0 \sim 280kG$ 
for the resonant $\Delta m^2_{01}$ can be excluded at more than $4\sigma$. A low rate would 
signal new physics, excluding all the 90$\%$ allowed range of the standard LMA solution at 3$\sigma$, 
and a time variability would be a strong signature of the RSFP model.
The CNO fluxes are the ones for which the two SSM predictions exhibit the largest
differences, so their measurement at SNO+ will be important to favour one or the other.
The distinction will be clearer after LMA or RSFP are confirmed with pep, but still,
a CNO measurement at the level of SSM~I/LMA will disfavour SSM~II at about $3\,\sigma$.
We conclude that consistency between future pep and CNO flux measurements at SNO+ and 
Chlorine would either favour an LMA+RSFP scenario or favour SSM~II over SSM~I.
\end{abstract}
\end{titlepage}   

\renewcommand{\thefootnote}{\arabic{footnote}}
\setcounter{footnote}{0}
\section{Introduction} 

Neutrino oscillations in matter \cite{Wolfenstein:1977ue} with the resonant 
amplification of a small vacuum mixing angle \cite{Mikheev:1986wj},
although a much attractive mechanism, has not proven to be the origin of 
the solar neutrino deficit. While also an oscillation and resonant effect,
the large mixing angle solution (LMA) \cite{Vignaud,Bahcall:1998jt,Bahcall:2003ce} 
has instead become generally accepted as the dominant one
\cite{Eguchi:2002dm,Araki:2004mb,Aharmim:2005gt}. In the LMA mechanism the 
conversion from active electron neutrinos produced in the solar core to weakly 
interacting ones of another flavour takes place through a strongly adiabatic 
resonance occuring still at the solar core. The order parameter is a large vacuum 
mixing of the order of $30^{o}-33^{o}$.

On the other hand, the time variability of the active neutrino event rate had been 
hinted long ago by the Chlorine collaboration \cite{Homestake} who suggested a 
possible anticorrelation of active neutrino flux with sunspot activity. It was 
then interpreted by Voloshin, Vysotskii 
and Okun \cite{VVO} as a neutrino magnetic moment effect, such that an intense sunspot
activity would induce a conversion of a large fraction of neutrinos into undetectable
ones (either steriles or of a different flavour) through the interaction of the 
magnetic moment with the solar magnetic field. Hence
a more intense solar activity would correspond to a smaller flux of detectable neutrinos
and viceversa. An interesting evolution of this proposal was the suggestion in 1987 by
Lim and Marciano \cite{RSFP1} and by Akhmedov \cite{RSFP2} that the neutrino spin
flavour precession could take place anywhere inside the sun via a resonant process.
This enhances the mechanism and allows for a smaller neutrino magnetic moment in 
order to produce a visible effect. The resonant spin flavour precession (RSFP)
bears a resemblance to matter oscillations and is a result of the balance 
between matter density and the product $\mu_{\nu}B$ (neutrino magnetic moment times 
the solar field).

The interpretation of solar data is at present still partly ambiguous, with several 
scenarios involving RSFP \cite{Chauhan:2005pn,Barranco:2002te} and non-standard neutrino
interactions \cite{Guzzo:2001mi} being viable.
Our knowledge of the solar neutrinos relies essentially on the data from the high
energy sector (mainly the $^8 B$ flux), whereas the overwhelming low and intermediate
energy one remains vastly unknown, except for the integrated measurements provided
by the radiochemical experiments. As pointed out earlier \cite{Chauhan:2004sf},
the observed decrease of the Gallium event rate \cite{Cattadori} opens the question
of whether there is time variability affecting only the low energy sector and has 
motivated the investigation of RSFP to light sterile neutrinos in
combination with LMA \footnote{The magnetic moment we refer here is of course a 
Majorana transition one since it connects active neutrinos to sterile ones, hence
of a different flavour.}. LMA+RSFP thus requires a sizable neutrino 
magnetic moment and a strong field at times of intense solar activity and a weak 
field otherwise, which causes the modulations in the neutrino event rate.
Low energy solar neutrino experiments like Borexino \cite{Arpesella:2001iz,
Collaboration:2007xf}, KamLAND \cite{Eguchi:2002dm,Araki:2004mb} and SNO+ will 
no doubt help in clarifying the situation within the next few years. 
The first Borexino data, very recently released \cite{Collaboration:2007xf}, are 
compatible with the RSFP model predictions \cite{Chauhan:2005pn,Chauhan:2004sf} 
earlier derived since those data were taken during the present year (2007) when the 
magnetic solar activity is at a minimum.

Besides our limited knowledge of the low energy neutrino sector, one of the key 
inputs of solar models, namely the amount of heavy element abundance relative to 
hydrogen, $Z/X$, is still unclear \cite{Basu:2006vh}. This affects the neutrino
fluxes, in particular most strongly the CNO ones and to a lesser extent the 
$^7 Be$ one. Two classes of standard solar models (SSMs) may at present be 
distinguished \cite{Bahcall:2005va,Serenelli:2007zz}, one with a 'high' value 
of $Z/X$ \cite{GS98} and another with a 'low' one \cite{AGS05}, hereby denoted 
by SSM~I and SSM~II, respectively. In the first the metallicity is 
consistent with sound speed, convective zone depth and density profiles, in 
excellent agreement with helioseismology. Such is not the case in the second one 
which is however based on an improved modeling of the solar atmosphere. 
As uncertainties are considerable, we will in the present paper take both 
models into account \footnote{It should be mentioned however that SSM~I is 
referred to as the 'preferred' solar model \cite{Bahcall:2005va}.}.

The purpose of this paper is to analyse the possibility
of the proposed SNO+ experiment to ascertain on whether RSFP occurs at low 
and/or intermediate energies.
We will also investigate the possible signatures of the two 
classes of SSMs, in particular whether it is possible to distinguish between
them with data from the forthcoming SNO+ and from the Chlorine 
\cite{Cleveland:1998nv} experiment. 
For neutrino-electron scattering experiments, sensitivity to neutrino physics
depends on the accuracy of solar models, since the calculation of the
electron neutrino survival probability relies on the comparison of the
measured flux with the total predicted flux. So for the purpose of
distinguishing models with different survival probabilities, the best
sensitivity lies in the observation of the solar flux component with the
smallest error. Above the threshold of liquid scintillator electron
scattering experiments, the component with the most accurate prediction is
the monoenergetic (1.442 MeV) source of pep neutrinos. The pep flux also has the 
smallest spread in the predictions from the
main solar models, which allows its measurement to be sensitive to Neutrino
physics without needing to choose a particular solar model.

The paper is divided as follows: in Section {\bf \ref{sec:2}} we briefly review the RSFP
model and its motivations, and compare its predictions with standard LMA for the 
existing Chlorine experiment data, in the context of the two SSMs.
In Section {\bf \ref{sec:SNO}} we briefly describe the SNO+ experiment. 
In Sec. {\bf \ref{sec:rates}} we describe the method and present our results.
In Section {\bf \ref{sec:analysispep}} we comment on the 
sensitivity of the pep measurement to an RSFP effect and we
investigate the possibility for SNO+ to distinguish between SSMs of the two types
with the CNO measurement in Sec.  {\bf \ref{sec:analysiscno}}.
Finally in Section {\bf \ref{sec:conclusions}} we report our conclusions.

\section{Resonant spin flavour precession and solar models}\label{sec:2}
In this Section we describe the RSFP effect from active to light sterile
neutrinos and evaluate the survival probability for LMA and LMA+RSFP (non-vanishing 
$\mu_{\nu}$ with low and high solar magnetic field). We start with a brief 
review of the model presented in ref. \cite{Chauhan:2004sf} whose predictions 
for Borexino and LENS were analyzed in \cite{Chauhan:2005pn} and for 
KamLAND in \cite{Chauhan:2005ju}. We also evaluate the event rates in the Chlorine 
experiment for standard neutrinos (i.e. massless and with no magnetic moment), 
for the LMA and LMA+RSFP scenario and for standard LMA (vanishing $\mu_{\nu}$). 
We use the neutrino fluxes from the type I \cite{GS98} and II \cite{AGS05} solar models.

\subsection{Resonant Spin Flavour Precession}\label{sec:RSFP}

The possible anticorrelation with sunspot activity of the electron neutrino flux 
in the Ga experiments is most naturally explained in terms of a 
resonant conversion to neutrinos of other types that are unseen by the weak charged 
current. As shall be discussed, this requires a mass square difference between 
the intervening neutrino flavours in the resonance $\Delta m^2_{01}=O(10^{-8}eV^2)$.
Such a value implies that conversion to weakly interacting neutrinos is excluded,
leaving us the possibility of conversion to sterile neutrinos. The simplest departure 
from conventional LMA able to generate such a conversion is provided by a model 
which, in addition to the two flavours involved in LMA, introduces a sterile neutrino
with a vanishing vacuum mixing. The active states $\nu_e,\nu_{\mu}$ communicate 
to the sterile one via a single magnetic moment. Owing to the large order of 
magnitude difference between the parameters $\Delta m^2_{21}$ and $\Delta m^2_{01}$ 
(associated with the LMA and SFP resonances respectively) 
the two resonances are located far apart, so that they do not interfere. 
A straightforward but long calculation leads to the following form of the Hamiltonian 
\cite{Chauhan:2004sf} 
\begin{equation}
\cal{H}_{\rm {M}}=\left(\begin{array}{ccc}\frac{-\Delta m^2_{10}}{2E}&
\mu_{\nu}B& 0 \\ \mu_{\nu}B& \frac{\Delta m^2_{21}}{2E}s^2_{\theta}+V_e&
\frac{\Delta m^2_{21}}{4E}s_{2\theta}\\ 0 &\frac{\Delta m^2_{21}}{4E}s_{2\theta}&
\frac{\Delta m^2_{21}}{2E}c^2_{\theta}+V_{\mu}\end{array}\right)
\end{equation}
in the mass matter basis $(\tilde\nu_0~\tilde\nu_1~\tilde\nu_2)$. 
In this equation $V_e$, $V_{x}$ are the matter induced potentials for
$\nu_e$ and $\nu_{x}$, $B$ is the solar magnetic field and $\theta$ is the 
vacuum mixing angle. 

The important transition with order parameter $\mu_{\nu}B$ and whose time dependent 
efficiency may determine the possible modulation of neutrino flux is therefore
between mass matter eigenstates $\tilde\nu_0$, $\tilde\nu_1$. It is expected
to resonate in the region where the magnetic field is the strongest in the
period of high solar activity.
We consider the Landau Zener approximation in dealing with the two resonances. 
Since the LMA one is strongly adiabatic, we need only to consider the 
jump probability between $\tilde\nu_0$, $\tilde\nu_1$
in the vicinity of the SFP resonance \footnote{For calculational details 
see ref. \cite{Chauhan:2004sf}.}.

In the LMA+RSFP scenario active neutrinos are
partially converted to light sterile ones at times of strong solar magnetic 
field, thus leading to the lower Gallium event rate in the period 1998-2003 
\cite{Cattadori}, while LMA acts alone otherwise and the higher rate is obtained. 
As in previous publications, throughout our RSFP calculations 
we use a value for the neutrino magnetic moment $\mu_{\nu}=10^{-12}\mu_B$.
For a definite neutrino energy the critical density is fixed by the order of magnitude
of the corresponding mass squared difference and determines the resonance range. 
Furthermore it has been noticed \cite{Caldwell:2005hp} that the solar rotation 
frequency matches the observed neutrino modulation in the equatorial section of 
the convection zone near the tachocline, at $\simeq(0.7-0.8)R_S$. This indicates
a field profile whose time dependent peak occurs around this depth. 
In order that the low and intermediate energy neutrino SFP resonance is located
in this range, so as to provide time modulation in this sector, one needs
$\Delta m^2_{01}= O(10^{-8})eV^2$. Hence, owing to the absence of interference
between the two resonances (LMA and RSFP) the high energy solar neutrino experiments 
(SuperKamiokande \cite{Fukuda:2002pe} and SNO \cite{Aharmim:2005gt}) are not 
expected to exhibit any time modulation in their
event rate. On the contrary, of all the experiments so far, Gallium will be the most 
sensitive of all to variability, while Chlorine may lie in the borderline with some
moderate variability which failed to be clearly detected.

Based on the above criteria we chose solar field profiles as in fig.1 of ref.
\cite{Chauhan:2005pn}. In our previous publications \cite{Chauhan:2004sf,Chauhan:2005pn,
Chauhan:2006yd} we performed two separate fits for the high and low Gallium event rates
together with all other solar neutrino data. We also investigated the dependence
of the results on the choice of solar field profile and on the value of $\Delta m^2_{01}$.
We will in the present paper use our best choices from ref. \cite{Chauhan:2006yd}, namely

\begin{equation}
B=\frac{B_0}{cosh[6(x-x_{c})]}~~~0<x<x_{c}
\end{equation}
\begin{equation}
B=\frac{B_0}{cosh[15(x-x_{c})]}~~~x_{c}<x<1
\end{equation}
for the field profile and   
$$\Delta m^2_{01}=-1.7\times 10^{-8}eV^2.$$
Here $B_0$ is the peak field value, which for active sun we take to be $280kG$, $x$ is the 
fraction of the solar radius and $x_c=0.71$.

\begin{figure}[tb]
\centering
{\epsfig{file=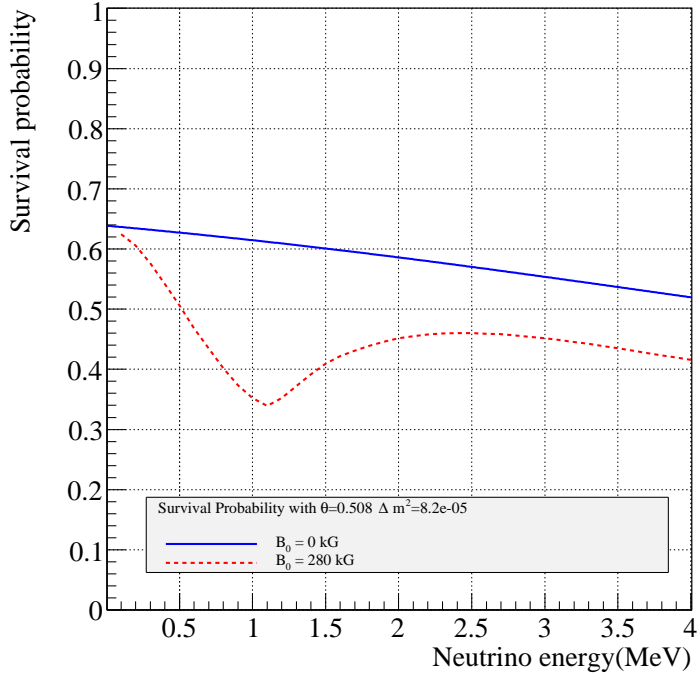,height=10.cm}}
\caption{Electron neutrino survival probability for LMA (zero field) and
         LMA+RSFP for the considered point. For the RSFP case we use a peak field 
         value $B_0=280kG$ and a profile as in
         \cite{Pulido:2006yn,Chauhan:2006yd}.}
\label{fig:survprob}
\end{figure}

Furthermore we use
\cite{Pulido:2006yn,Chauhan:2006yd} 
\begin{equation}\label{eq:LMASFP}
\theta=0.508,~~\Delta m^2=8.2\times 10^{-5}eV^2
\end{equation}
which lie within $1\sigma$ of the central values for KamLAND\cite{Araki:2004mb}.
The field profile chosen, together with the value of $\Delta m^2_{01}$, will be
responsible for a modulation shared by $pp$ and $^7 Be$ neutrinos and a dip at
the low and intermediate energy in the survival probability.
This is shown in fig.\ref{fig:survprob}.

\subsection{Chlorine rate, type I and II solar models}\label{sec:situation}
\begin{center}
\begin{table}[tb]
\begin{tabular}{||l|c|c||c|c||}
\hline\hline
 & \multicolumn{2}{|c||}{$\phi_{j} (cm^{-2}s^{-1})$} & \multicolumn{2}{|c||}{$R_{j}$ (SNU)} \\
\hline & SSM~I & SSM~II & SSM~I & SSM~II \\
\hline\hline
\textit{pep} & $1.42\times10^{8} (1\pm0.015)$          & $1.45\times10^{8} (1\pm0.01)$             & 0.222 $\pm$0.003  & 0.226 $\pm$0.002 \\
$^{7} Be$     & $4.84\times10^{9} (1\pm0.105)$          & $4.34\times10^{9} (1\pm0.093)$            & 1.16 $\pm$0.122 & 1.043 $\pm$0.097 \\
$^{8} B$      & $5.69\times10^{6} (1\pm0.16)$           & $4.51\times10^{6} (1\pm0.12)$             & 6.740 $\pm$1.078  & 5.342 $\pm$0.641 \\
$^{13} N$     & $3.05\times10^{8} (1\pm^{0.31}_{0.28})$ & $2.00\times10^{8} (1\pm^{0.145}_{0.127})$ & 0.052 $\pm^{0.016}_{0.014}$ & 0.034 $\pm^{0.005}_{0.004}$ \\
$^{15} O$     & $2.31\times10^{8} (1\pm^{0.33}_{0.29})$ & $1.44\times10^{8} (1\pm^{0.165}_{0.142})$ & 0.154 $\pm^{0.051}_{0.045}$ & 0.096 $\pm^{0.016}_{0.014}$ \\
\hline
\multicolumn{3}{||r||}{$\Sigma R_{Cl}$} &  $8.33\pm1.2$1 & $6.74\pm0.73$ \\
\hline\hline
\end{tabular}
\caption{Predicted total fluxes ($\phi_{j}$) and expected event rates ($R_{j}$) in the Chlorine experiment for standard neutrinos.}
\label{tab:normalization}
\end{table}
\end{center}

Here we analyse the Chlorine event rates in the
two scenarios, the one with and the one without neutrino magnetic moment,
using the neutrino fluxes from SSM~I \cite{GS98} and SSM~II \cite{AGS05}. 
It will be seen, from the comparison of these predictions with the Chlorine 
data \cite{Cleveland:1998nv} 
\begin{equation}\label{eq:RCl}
R_{Cl}=2.56\pm0.16\pm0.15~SNU,
\end{equation}
that RSFP is compatible with both solar models while standard LMA favours SSM~II. 
The predicted fluxes $\phi_j$, 
the partial event rates $R_j$ for each flux and the total event rate for the 
Chlorine experiment $R_{Cl}$ are given in table ({\bf\ref{tab:normalization}})
for both models with standard
neutrinos. The errors in the partial event rates $\delta R_{j}$ listed in this 
table were calculated from
\begin{equation}
R_{j}=\int\, \sigma{_{_{Cl}}}\,\phi{_{_j}}
	\left(1\pm \frac{\delta \phi_{j}}{\phi_{j}}\right)dE_\nu
	=\bar R_{j} \pm \delta R_{j}
\end{equation}
for flux $j$ that is, only the flux errors were considered.

\begin{table}[tb]
\begin{center}
\begin{tabular}{||l|c|c|c|c||}
\hline\hline
 &  \multicolumn{2}{|c|}{$R_{LMA}$ (SNU)} & \multicolumn{2}{|c||}{$R_{RSFP}$ (SNU)} \\
\hline & SSM~I & SSM~II & SSM~I & SSM~II \\
\hline\hline
\textit{pep}    & 0.133 $\pm$0.002            & 0.136 $\pm$0.001            & 0.088 $\pm$0.001            & 0.090 $\pm$0.001 \\
$^{7} Be$        & 0.710 $\pm$0.075            & 0.637 $\pm$0.059            & 0.447 $\pm$0.047            & 0.401 $\pm$0.037 \\
$^{8} B$         & 2.003 $\pm$0.320            & 1.587 $\pm$0.191            & 1.888 $\pm$0.302            & 1.497 $\pm$0.180 \\
$^{13} N$        & 0.032 $\pm^{0.010}_{0.009}$ & 0.021 $\pm{0.003}$ & 0.019 $\pm^{0.006}_{0.005}$ & 0.012 $\pm{0.002}$ \\
$^{15} O$        & 0.091 $\pm^{0.030}_{0.026}$ & 0.057 $\pm^{0.009}_{0.008}$ & 0.059 $\pm^{0.019}_{0.017}$ & 0.037 $\pm^{0.006}_{0.005}$ \\
\hline
\multicolumn{1}{||r|}{$\Sigma R_{Cl}$} & 2.97$\pm$0.40 & 2.44$\pm$0.25 & 2.50$\pm$0.35 & 2.04$\pm$0.22 \\
\hline\hline
\end{tabular}
\caption{Expected event rates in the Chlorine experiment for the two SSMs. For the parameter choices
used in the LMA and LMA+RSFP cases, see the main text [eq. (\ref{eq:LMASFP})] and fig. \ref{fig:survprob}.}
\label{tab:clevents}
\end{center}
\end{table}

The results for the event rates are given in table ({\bf\ref{tab:clevents}}) for LMA with 
parameters as in eq. (\ref{eq:LMASFP}) and LMA+RSFP.
The errors in the total Chlorine rate are obtained using the correlation among the
flux errors as in tables 16 and 17 of ref. \cite{Bahcall:2005va}.
Comparing the event rates for LMA (zero field) and LMA + RSFP (high field), it is 
seen that some modulation is expected in both SSMs, due to the time dependence
of the solar magnetic field. 

Averaging the results 
\begin{equation}
SSM~I~~~~\bar R_{Cl}=2.73\pm0.38SNU,
\end{equation}
\begin{equation}
SSM~II~~~~\bar R_{Cl}=2.24\pm0.23SNU
\end{equation}
we see that both SSMs are fully compatible with the data. 
That is not the case when using for $\theta_{12},~\Delta m^2_{21}$ 
the central values of standard LMA,
that correspond to a negligible magnetic moment,
from global solar analysis and KamLAND
\cite{Aharmim:2005gt,Aliani:2003ns}:
\begin{equation}
\theta=0.592,~~\Delta m^2=8.0\times 10^{-5}eV^2.
\label{eq:fullLMA}
\end{equation}
In fact, in this case the Chlorine rate is fully 
compatible with the SSM~II model while SSM~I is slightly disfavoured at 
1.3$\sigma$. The predictions obtained 
from the solar neutrino and KamLAND global analysis
 central values are shown in Table (\ref{tbl:cllma}).

\begin{table}[ht]
\begin{center}
\begin{tabular}{||l|c|c||}
\hline\hline
 &  \multicolumn{2}{|c||}{$R$ (SNU)} \\
\hline & SSM~I & SSM~II \\
\hline\hline
\textit{pep}    & 0.120 $\pm 0.002$ & 0.123  $\pm 0.001$ \\
$^{7}Be$        & 0.638 $\pm 0.067$ & 0.572  $\pm 0.053$ \\
$^{8}B$         & 2.351 $\pm 0.376$ & 1.864  $\pm 0.224$ \\
$^{13}N$        & 0.028 $\pm ^{0.009}_{0.008}$ & 0.019 $\pm ^{0.003}_{0.008}$  \\
$^{15}O$        & 0.082 $\pm ^{0.027}_{0.024}$ & 0.051 $\pm ^{0.002}_{0.007}$ \\
\hline
\multicolumn{1}{||r|}{$\sum R_{Cl}$} & 3.22 $\pm 0.45$ & 2.63 $\pm0.27$ \\
\hline\hline
\end{tabular}
\caption{Expected event rates in Chlorine for standard LMA with Solar+KamLAND bestfit parameters: $\theta=0.592$, $\Delta m^{2}=8.0\times10^{-5}$.}
\label{tbl:cllma}
\end{center}
\end{table}

\section{The SNO+ experiment}\label{sec:SNO}
Among the existing and proposed solar neutrino experiments to come online
in the near future, SNO+ will be the only one with the ability to
measure a survival probability, as the ratio between the measured and 
SSM predicted rate of a solar neutrino component,  at the precision level of 5\%.
Due to the depth of SNOLAB, SNO+ will have a low level of $^{11}C$ background,
allowing for an accurate measurement of the the pep neutrino flux -
predicted in the SSM with an error of (1-1.5)\%. This is an advantage over the
measurement of the $^7Be$ flux, that can be done earlier in other
experiments (Borexino, KamLAND), but is predicted with an error of (9.3-10.5)\%.
In addition, SNO+ will also be the best experiment to measure the
CNO-cycle fluxes. 
Measurements of the pep and CNO fluxes will be possible also in Borexino\cite{borexpep},
although with larger systematic and statistical uncertainties than at SNO+.
The expected accuracy on these fluxes will allow for
the distinction between different Solar Models accepted by the present
data.
In addition, SNO+ will measure reactor anti-neutrinos, geo-neutrinos
and can later be upgraded to detect neutrinoless double-beta
decays, in order to search for the absolute neutrino mass.

\subsection{The SNO+ detector}
SNO+ \cite{SNO+,SNO+2} is a proposed upgrade of the the Sudbury Neutrino
Observatory (SNO) detector \cite{SNO}, in which the heavy water
target will be replaced by an organic liquid scintillator.
The scintillator was chosen to have a good light yield and
transparency, and to be compatible with the existing SNO
components: chemical and optical compatibility with the acrylic
vessel and emission wavelength peaks close to the PMT response.

SNO+ has a fiducial volume of ~1000 tonnes, in a 12m diameter
acrylic vessel viewed by 9456 PMTs mounted on an 18 m diameter geodesic
structure. The PMTs have a diameter of 20 cm, and are coupled to
optical reflectors, increasing the total effective coverage to 54\%.
The region outside the acrylic vessel is filled with 7000 tonnes of
light water (1700 tonnes within the PMT structure), acting as a shield
against external radiation. The external regions  viewed by additional
114 PMTs, to act as a veto for cosmic ray muons.

The large volume of the detector will allow for an effective veto of
external backgrounds (mainly gammas and neutrons) through position
reconstruction, while maintaining a large fiducial volume.
Concerning internal radioactivity, the purity levels achievable
for scintillators can be estimated from the existing KamLAND ones
for the U and Th chains, while others such as $^{40}K$ and $^{210}Bi$ are
being studied for several experiments and expected to be reduced
by several orders of magnitude.

In addition, the center of the SNO detector is at a depth of 2092m,
or 6010 m of water equivalent - it is located in the deepest
underground physics facility, SNOLAB.
The $^{11}C$ contamination produced by cosmic ray muons - which in
general prevents or severely hinders the measurement of the pep
solar neutrino line - is not a problem at SNO+ since at this depth
there are only approximately 70 muons to enter the detector per day,
and so the data taken just after can be cut away.

The fact that SNO+ is located in a laboratory whose background
conditions are well known and has the same geometry as the SNO
detector, allows for accurate estimations of its capabilities even
before construction.

\subsection{Neutrino Measurements at SNO+}
Neutrinos interact in the scintillator through elastic scattering off
electrons ($\nu e^- \to \nu e^-$). In the energy range of pep solar
neutrinos, the cross-section is around five times
smaller for muon or tau neutrinos than for electron neutrinos, so SNO+ is
primarily a "disappearance experiment" for electron neutrinos, even
if there is some sensitivity to the other flavors of active neutrinos.

The high light yield of the scintillator - about 100 times
more light than the Cerenkov light in heavy water - allows the detection
of recoil electrons with energies as low as tens of keV.

Since the measured energy will be roughly proportional to the number of 
detected scintillation photons, the main contribution to the energy 
resolution can be written as:
\begin{equation}\label{eq:resolution}
\frac{\sigma(E)}{E}=\frac{1}{\sqrt{N_{ph}}} \times \frac{1}{\sqrt{E}}
\end{equation}
where $N_{ph}$ is the detected light yield which, from conservative Monte 
Carlo simulations \cite{SNO+3},  was estimated at about $N_{ph}=600$ 
photons per MeV.
This results in an energy resolution of 4\% at 1 MeV, better than in 
other large liquid scintillator detectors -- 6.2\% at KamLAND\cite{Araki:2004mb} 
and 5\% at Borexino\cite{Borexino}.

The energy of the incoming neutrino is not directly reconstructed,
but the different electron spectra structures - and mainly the Compton
edge directly related to the neutrino energy - can be used to 
statistically separate the neutrino signal from background and the different
contributions to the solar neutrino flux, namely the mono-energetic
pep line.

The high rate low energy background from $^{14}C$ in the scintillator and
the gamma background from detector materials can be suppressed with data
selection cuts in both energy -- threshold of 500 keV -- and position. 
For the pep analysis, an energy window from 800 keV to 1500 keV is expected,
to avoid background from fluctuations of the $^7Be$ signal on the lower side.

Assuming U and Th background levels from KamLAND and the expected levels
for $^{40}K$ and $^{210}Bi$ after purification, SSM shapes for
CNO neutrinos and LMA oscillation, a likelihood fit to the energy spectrum
allows the separate measurement of the number of pep (CNO) neutrino events 
with a 4\% (6\%) uncertainty\cite{SNO+3}, 
after three years of data taking. Adding in a global systematic uncertainty 
on the fiducial volume, estimated at 3\%, a total flux measurement 
error of 5\% is obtained for pep.

\section{Sensitivity of SNO+ to RSFP and solar models}\label{sec:sensitivity}
In this section we calculate the predicted event rates of pep and CNO solar
neutrinos in SNO+ for the LMA+RSFP model in the high and low field cases, for
 the two sets of solar models, SSM~I and SSM~II (Sec. \ref{sec:rates}). 
We then use the predicted rates to estimate how sensitive is SNO+ to an RSFP effect
(Section \ref{sec:analysispep}),
and to the discrimination between the two solar models, 
SSM~I and SSM~II (Section \ref{sec:analysiscno}).

\begin{figure}[htb]
\centering
{\epsfig{file=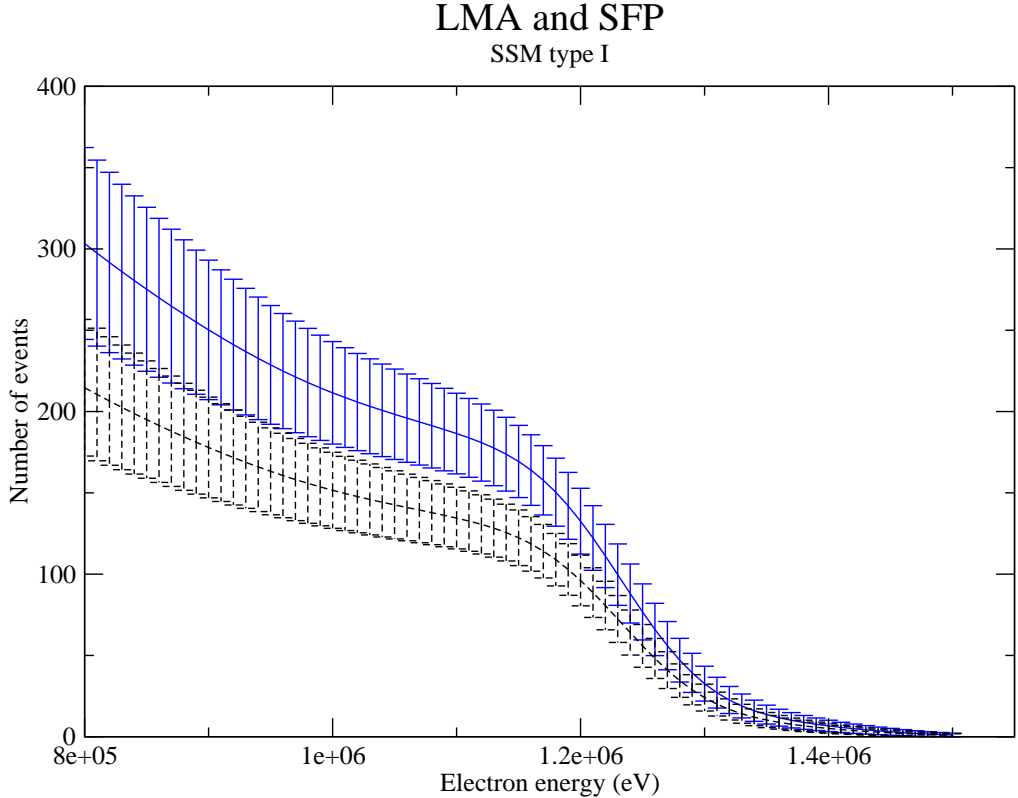,height=16.cm,angle=-90}}
\caption{Standard LMA and LMA+RSFP spectra in the active sun period
for type I model (SSM~I \cite{GS98}).}
\label{fig:spectrumSystISFP}
\end{figure}
\begin{figure}[htb]
\centering
{\epsfig{file=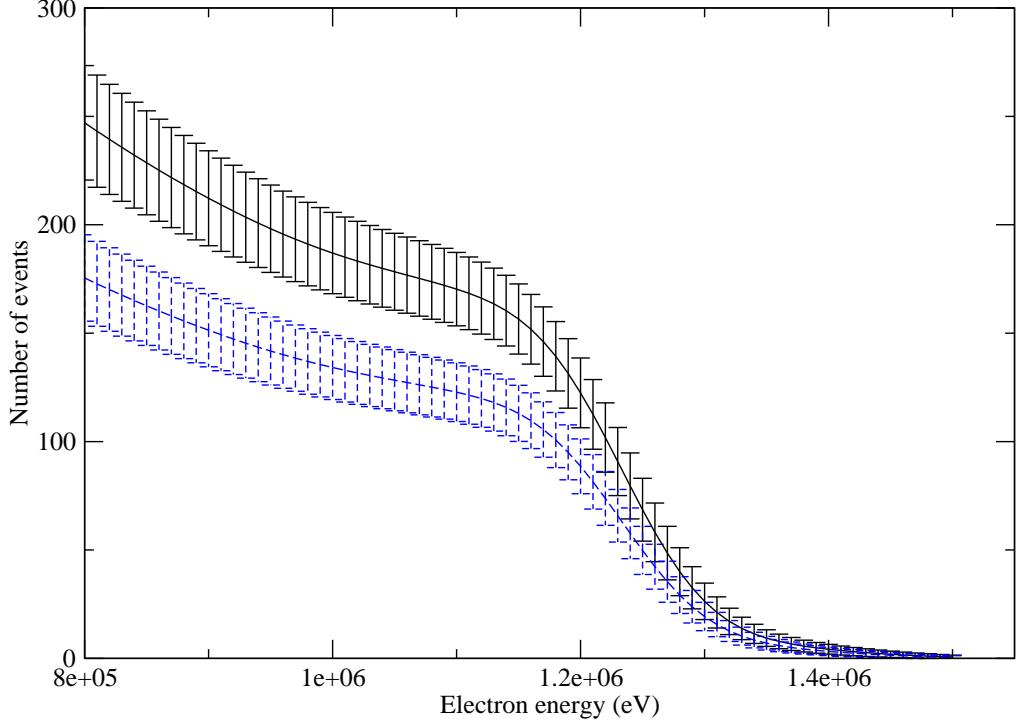,height=16.cm,angle=-90}}
\caption{Same as fig.3 for type II model (SSM~II \cite{AGS05}).}
\label{fig:spectrumSystIISFP}
\end{figure}
\begin{figure}[htb]
\centering
{\epsfig{file=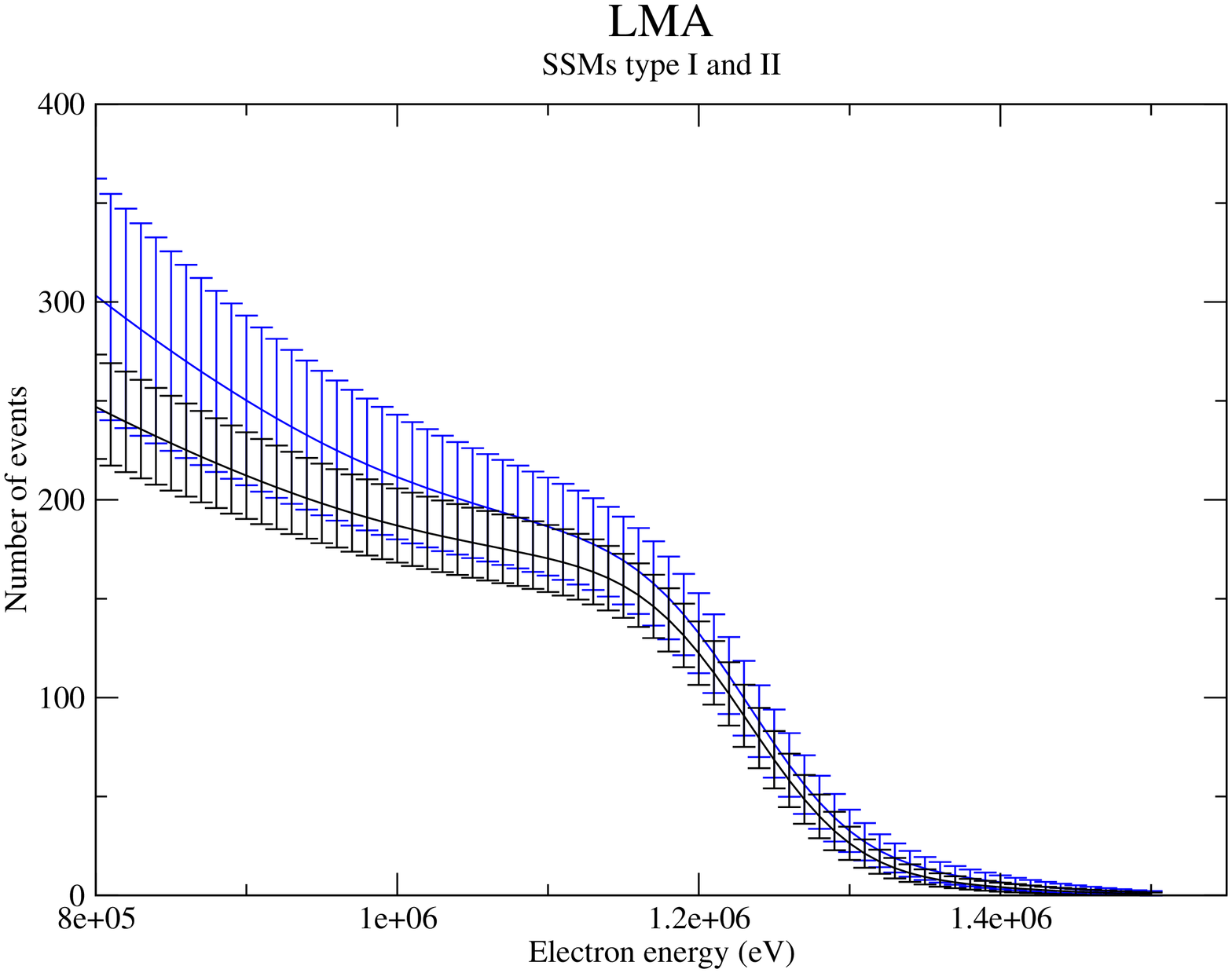,height=16.cm,angle=-90}}
\caption{Standard LMA spectrum for models of type I (SSM~I \cite{GS98})
and type II (SSM~II \cite{AGS05}).}
\label{fig:spectrumSystILMA}
\end{figure}

\subsection{Expected rates and spectral analysis}\label{sec:rates}
The expected event rate for each solar neutrino component can be given by

\begin{eqnarray}\label{eq:N}
R(j)&=&V_{fid}\,\rho_e\,\int_{T_{min}}^{T_{max}} dT_m
      \int_0^\infty dT\, F(T,T_m)
      \int_0^{E_{max}} dE\, \phi_j \Phi_j(E)\, P_{\nu_x}(E)\, \frac{d\sigma_{\nu_x}(E,T)}
      {dT}.
\end{eqnarray}
The total fluxes $\phi_j$ for each component $j\in\{pep,N,O\}$ are taken from table 
({\bf\ref{tab:normalization}}). $\Phi_j(E)$ is
the normalized neutrino flux spectrum for flux $j$ (delta function for pep).
Eq. (\ref{eq:N}) includes a sum over $x\in\{e,\mu,\tau\}$.
The quantity $T_m$ is the observed electron kinetic energy, 
$T$ is the true electron kinetic energy, $E$ is the neutrino energy with
$E_{max}$ given by the kinematics and $T_{min}$, $T_{max}$ are the lower and upper thresholds
of the $T_m$ analysis energy window, given in Section~\ref{sec:SNO}. $\rho_e$ is the electron 
density of the scintillator and $V_{fid}$ is the analysis fiducial volume.
$P_{\nu_x}(E)$ is the probability for $\nu_e$ $\rightarrow$ $\nu_x$ conversion,
$d\sigma_{\nu_x}(E,T)/dT$ is the cross section for $\nu_x e$ scattering and
$F(T,T_m)$ is the response function of the detector. This is given by
\begin{eqnarray}
F(T,T_m)&=&\frac{1}{\sigma\sqrt{2\pi}}\exp\left( -\frac{(T-T_m)^2}{2\sigma^2}\right)
\end{eqnarray}
where $\sigma$ is the detector energy resolution, as described by 
Eq.~\ref{eq:resolution}. 
For models I and II we compute the contribution
from each component of the solar neutrino spectrum:
$pep$, $^{15}O$, and $^{13}N$
(the $^{17}F$ contribution is negligible and is not evaluated).

We perform the rate calculation for four scenarios:
\begin{description}
\item[SFP]{The oscillation probabilities $P_{\nu_x}$ are
   computed with the RSFP effect in addition to LMA as explained in
Section {\bf\ref{sec:RSFP}} and in \cite{Chauhan:2005pn,Chauhan:2004sf}.
We consider a high field value of $B_0=280kG$, which would correspond to the first three
years of data taking, since SNO+ is expected 
to start during the next period of rising solar activity including its 
maximum around 2011\cite{solar}.}
\item[SFP0]{$P_{\nu_x}$ are obtained from the LMA oscillation effects
   with $\Delta m^2,~\theta$ as in eq. (\ref{eq:LMASFP}), corresponding to $B_0=0kG$, 
   the RSFP prediction for low solar activity periods.}
\item[LMA]{$P_{\nu_x}$ are obtained from the LMA oscillation effects
   with Solar+KamLAND bestfit $\Delta m^2,~\theta$, as in eq. (\ref{eq:fullLMA}).}
\item[NOsc]{For reference we also evaluated the case for the absence of oscillations,
  where $P_{\nu_x}=\delta_{x,e}$\ .}
\end{description}

The extraction of the pep and CNO signals from the future experimental data will be based 
on a fit to the measured energy spectrum, which will require a very accurate knowledge of the
detector response, as well as of the residual backgrounds that pass the selection cuts, based on 
extensive detector calibrations. The SNO+ collaboration has carried Monte Carlo simulations
\cite{SNO+3} of the expected backgrounds in the pep-CNO energy window, and performed maximum
likelihood signal extraction on the simulated signals and backgrounds, mostly from isotopes
of the $^{238}U$ and $^{232}Th$, as well as $^{40}K$. The resulting
sensitivity strongly depends on the background levels. Assuming the target values for
KamLAND, a sensitivity of 4\% for pep and 6\% for CNO is obtained for three years of data.

This calculation assumed standard LMA oscillations, and might be changed in the case of RSFP, 
in which the signals are reduced. We carry out a simple spectral analysis, without considering 
the backgrounds, for all cases. 

The energy spectrum, or expected number of events in the spectral bin i is given by:
\begin{eqnarray}\label{eq:Nj}
N_i&=&\epsilon\, \int_{T_i}^{T_{i+1}} dT_m
      \int_0^\infty dT\, F(T,T_m)
      \int_0^{E_{max}} dE\, \phi_j \Phi_j(E)\, P_{\nu_x}(E)\, \frac{d\sigma_{\nu_x}(E,T)}
      {dT}
\end{eqnarray}
where $\epsilon=Livetime\times V_{fid}\rho_e\,$ is the exposure, and the sum extends over 
the flux index $j\in\{pep,N,O\}$. 

The contributions to uncertainty ($\delta N_i$) include:

\begin{itemize}
\item the statistical error, considering the number of events N for 3 years;
\item the energy scale error (we assumed an error of $10 keV$ in the determination of the threshold);
\item a global systematic error of 3\% from the fiducial volume determination;
\item the error in the total flux predictions as given in table ({\bf\ref{tab:normalization}}).
\end{itemize}
The error for each bin is obtained by adding quadratically these four sources of error: 
\begin{eqnarray}
\delta N_i &=& \sqrt{  \Big(\sqrt{N_i}\Big)^2
                    +\Big(\delta N_i^{\mbox{scale}}\Big)^2
		    +\Big(N_i*3\%\Big)^2
                    +\Big(\delta N_i^{\mbox{flux}}\Big)^2},
\end{eqnarray}
where the first three terms are experimental errors
(statistical, energy scale, and global systematic uncertainty)
and the last is the theoretical one (flux uncertainty).
Our extraction uncertainties are in reasonable agreement with the values quoted in \cite{SNO+3}, 
before including the theoretical and energy scale errors, and can thus be taken as indicative values. 
As expected, the CNO extraction uncertainty is increased in the RSFP case, and that is taken into 
account by this procedure. 
Equation \ref{eq:Nj} can be rewritten in matrix form as:
\begin{eqnarray}\label{eq:NFa}
N_i = F_{ij}\; \phi_j.
\end{eqnarray}
Here the normalized predicted spectra of each component $F_{ij}$ is a $100\times 3$ matrix. 
Once the data on $N_i$ are known, one can extract $\phi_j$ by inverting F:
\begin{eqnarray}\label{eq:alpha}
\phi_j = F^{-1}_{ji}\; N_i
\end{eqnarray}
where the $3\times 100$ matrix $F^{-1}=\left(F^T\; F\right)^{-1}\; F^T$
is the pseudoinverse of $F$ and the transpose of $F$, $F^T$, 
is a $3\times 100$ matrix.
The errors are also calculated from this matrix inversion, assuming full correlation
(uncorrelation) between the systematic (statistic) errors in each bin.

The spectra used in this analysis are shown in Figures 
\ref{fig:spectrumSystISFP}-\ref{fig:spectrumSystILMA}. 
The rate results for type I and II models are shown 
in Tables ({\bf\ref{tab:globalI}}) and ({\bf\ref{tab:globalII}}).

\begin{table}[ht]
\begin{center}
\begin{tabular}{||l|r|r|r|r||}
\hline\hline
 &\multicolumn{4} {c||}{Number of events (in thousands) from SSM~I}\\
\hline
Component &\multicolumn{1} {c|}{RSFP} &\multicolumn{1} {c|}{RSFP0} &\multicolumn{1}{c|}{LMA} &\multicolumn{1}{c||}{NOsc}\\
\hline\hline
\textit{pep} & $4.82\pm0.28$    & $6.55\pm0.38$   & $6.09\pm0.36$   & $9.98\pm0.58$\\
             &\small{$(\pm0.11)$}&\small{$(\pm0.15)$}&\small{$(\pm0.14)$}&\small{$(\pm0.23)$}\\
\hline
$^{13}N$     & $0.26\pm0.09$    & $0.40\pm0.14$   & $0.38\pm0.13$   & $0.60\pm0.21$\\
             &\small{$(\pm0.03)$} &\small{$(\pm0.05)$}&\small{$(\pm0.05)$}&\small{$(\pm0.07)$}\\
\hline
$^{15}O$     & $2.61\pm0.86$    & $3.63\pm1.20$   & $3.38\pm1.11$   & $5.52\pm1.82$\\
             &\small{$(\pm0.09)$} &\small{$(\pm0.13)$}&\small{$(\pm0.12)$}&\small{$(\pm0.20)$}\\
\hline
Total        & $7.69\pm1.04$    &$10.58\pm1.45$   & $9.85\pm1.35$   &$16.09\pm2.20$\\
             &\small{$(\pm0.23)$} &\small{$(\pm0.33)$}&\small{$(\pm0.31)$}&\small{$(\pm0.50)$}\\
\hline\hline
\end{tabular}
\end{center}

\caption{\small Expected number of events in 3 years of SNO+ from $pep$, $^{15}O$ and $^{13}N$ 
solar neutrinos for three years of data taking, assuming the type I model, considering 
LMA+RSFP, at high field (RSFP) and at low field (RSFP0), standard LMA and the no oscillation 
case.
The error in parentheses is from the energy scale uncertainty only. The total
error is indicated next to the predicted value and includes also the uncertainties
from the model, from statistics and systematics, as stated in the text.}
\label{tab:globalI}
\end{table}
\begin{table}[ht]
\begin{center}
\begin{tabular}{||l|r|r|r|r||}
\hline\hline
 &\multicolumn{4} {c||}{Number of events (in thousands) from SSM~II}\\
\hline
Component &\multicolumn{1} {c|}{RSFP} &\multicolumn{1} {c|}{RSFP0} &\multicolumn{1}{c|}{LMA} &\multicolumn{1}{c||}{NOsc}\\
\hline\hline
\textit{pep} & $4.92\pm0.29$    & $6.68\pm0.38$   & $6.22\pm0.36$   &$10.19\pm0.58$\\
             &\small{$(\pm0.11)$} &\small{$(\pm0.16)$}&\small{$(\pm0.15)$}&\small{$(\pm0.24)$}\\
\hline
$^{13}N$     & $0.17\pm0.03$    & $0.26\pm0.05$   & $0.25\pm0.05$   & $0.39\pm0.08$\\
             &\small{$(\pm0.02)$} &\small{$(\pm0.03)$}&\small{$(\pm0.03)$}&\small{$(\pm0.05)$}\\
\hline
$^{15}O$     & $1.63\pm0.27$    & $2.26\pm0.38$   & $2.11\pm0.35$   & $3.44\pm0.57$\\
             &\small{$(\pm0.06)$} &\small{$(\pm0.08)$}&\small{$(\pm0.08)$}&\small{$(\pm0.12)$}\\
\hline
Total        & $6.72\pm0.48$    & $9.25\pm0.67$   & $8.58\pm0.62$   &$14.02\pm1.01$\\
             &\small{$(\pm0.19)$} &\small{$(\pm0.27)$}&\small{$(\pm0.25)$}&\small{$(\pm0.41)$}\\
\hline\hline
\end{tabular}
\end{center}

\caption{\small Same as Table \ref{tab:globalI} for SSM~II.}
\label{tab:globalII}
\end{table}

\subsection{Analysis of the pep results}\label{sec:analysispep}

The most relevant aspect of the pep results is the large difference between the LMA and RSFP
predictions, that is not significantly affected by our present uncertainty on the solar model. 

In fact, even if the the lowest central value for the best-fit LMA prediction (in SSM~I) is 
measured, it is high enough to exclude the highest RSFP value (in SSM~II), at more than 4$\sigma$,
and if it would be down by 1$\sigma$ of the prediction, RSFP would still be disfavoured at
around 3$\sigma$. 

On the other hand, if SNO+ finds a pep flux lower than $1.40\times10^8\,cm^{-2}s^{-1}$,
it will necessarily follow that new physics is at work.
If the RSFP effect is observed, and the central value predicted in SSM~II
for $B_0 = 280 kG$ is measured, the best fit LMA point is excluded at 3.3$\sigma$. 
In this case, all the allowed $90\%$ C.L. region of LMA is also excluded
at the 3$\sigma$ level.

In addition to a low pep measurement, RSFP predicts significant time
variations of the measured flux with a solar cycle periodicity, due to the
dependence of the survival probability on the magnetic field peak.
This expected variation would allow
the distinction between RSFP and other scenarios that predict lower rates
regardless of the magnetic field.

\begin{figure}[htb]
\centering
{\epsfig{file=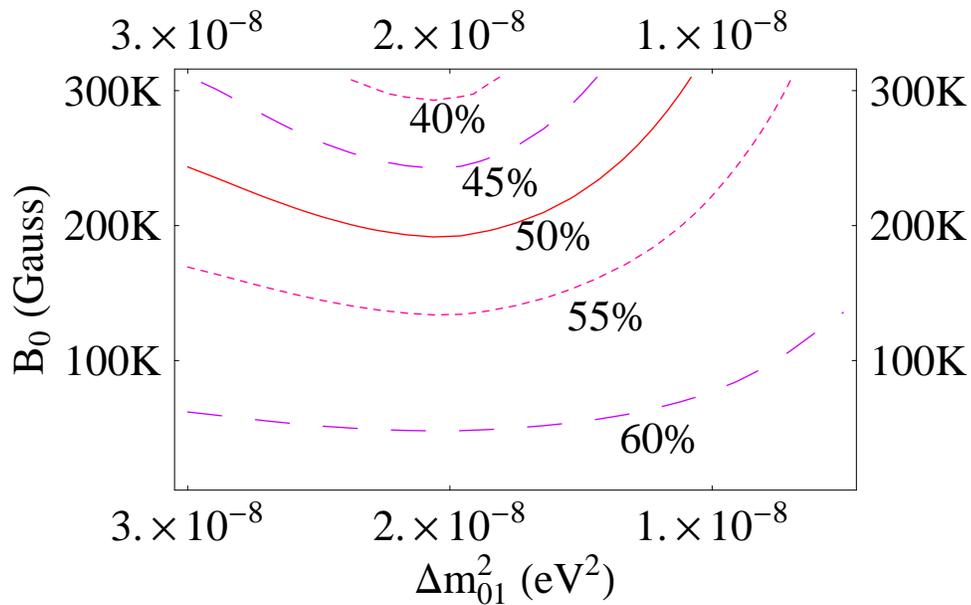,height=8.cm}}
\caption{The expected rate reduction for the $pep$ flux with respect to the non-oscillation case,
         as a function of the peak value $B_0$ of the solar magnetic field 
         and $\Delta m^2_{01}$.}
\label{fig:SFP}
\end{figure}

The effect is shown for field values different from $B_0 = 280 kG$ in figure \ref{fig:SFP},
for $\Delta m^2_{01}$ values around the resonant value obtained from
\cite{Pulido:2006yn,Chauhan:2006yd}.
We plot in fig \ref{fig:SFP} the expected rate reduction for the $pep$ flux 
in relation to the non-oscillation case
as a function of the solar magnetic field and $\Delta m^2_{01}$.
From tables ({\bf\ref{tab:globalI}}) and ({\bf\ref{tab:globalII}})
it is readily seen that $B_0=0$ (the x-axis in fig \ref{fig:SFP})
corresponds to a rate reduction of $65\%$.

\subsection{Analysis of the CNO results}\label{sec:analysiscno}

RSFP would change also the CNO rates in a consistent way, and that could serve 
as an independent cross-check of the oscillation model. However, the prediction
for CNO rates depends strongly on the SSM considered and has a large uncertainty 
within SSM~I.

In fact examining, for example, the $^{15}O$ contribution in the LMA case and
taking into account both the theoretical and experimental errors, we see that model I 
predicts $3383\pm1047$ events, while model II predicts of $2109\pm361$ events, 
so that the two results overlap only slightly. This is also the case for the RSFP 
predictions, so that after the oscillation pattern is established with pep, the
distinction between SSMs is improved.

If the central values of SSM~I would be measured, SSM~II could be excluded at 
around 2.6$\sigma$, both in LMA and RSFP cases. 
The $1\sigma$ upper edge of the prediction would allow a full exclusion at more
than $4.6\sigma$, while the lower edge is compatible with SSM~II.
The large uncertainty in SSM~I makes is harder to exclude,
even if the measurements are in full agreement with SSM~II predictions.
The results could, in turn, be used to constrain the upper values of 
CNO fluxes allowed in this model.

We see that SNO+ will be able to determine the $^{13}N$ and $^{15}O$ fluxes with 
such a precision that it can discriminate SSM of type I from SSM of type II, or
severely constrain SSM~I. 
There is no doubt that the discrimination of the several flux components with
this precision will have an impact on our present knowledge of the solar models.
However, an answer to the question {\em How well SNO+ discriminates SSM~I from 
SSM~II} can be obtained only after the experimental data are available.

\subsection {SNO+, Borexino and KamLAND}

We conclude this section with a brief comparison of the model predictions for
these three real time liquid scintillator experiments. We used the parameter 
values considered before ($\Delta m^2_{10}= 1.7\times 10^{-8}eV^2$, $B_0=280 kG$), 
the field profile as in eqs. (2), (3) and $\Delta m^2_{21}$, $\theta$ as in eq.(4). 
The results for the ratio between the predictions of RSFP and LMA with low field 
are shown in table ({\bf 6}) for Borexino and SNO+ (central values only). For 
KamLAND the same ratio for $Be$ gives 66.0\% (see ref.\cite{Chauhan:2005ju}).

We observe a sizable magnitude difference between $R_{Be}$ and $R_{pep}$. This  
is due to the fact that the $Be$ and $pep$ neutrino lines resonate at 0.68 and 0.78 
of the solar radius where the field is respectively 99\% and 61\% of its peak value.
All other results are consistent between Borexino and SNO+, taking the errors into 
account.

\begin{table}
\begin{center}
\begin{tabular}{|c|c|c|c||}
\hline\hline
Component  & Borexino & SNO+ \\
\hline
$R_{Be}$  & 65.7\% &       \\
$R_{pep}$ & 70.9\% & 73.6\% \\
$R_{N}$ & 66.7\% & 65.0\% \\
$R_{O}$ & 67.0\% & 71.9\% \\
\hline
\end{tabular}
\end{center}
\caption{\small Ratios between RSFP and LMA event rate predictions for Borexino 
(evaluated as in \cite{Chauhan:2005pn}) and for SNO+ (from tables 4 and 5).}
\end{table}

\section{Conclusions}\label{sec:conclusions}
One of the key questions that the SNO+ experiment will be able to address is 
the distinction between the two classes of SSMs, type I and II. 
Moreover SNO+ has the potential to distinguish the important issue of 
variability of the solar neutrino flux for low and intermediate energies.
In order to investigate the prospects for the SSM 
distinction we looked for an event rate prediction whose model dependence leads
to two well separated answers. 
The $^8$B total flux measured by SNO stands in between the SSM predictions and so
cannot resolve the ambiguity. Furthermore the Chlorine data can also not provide a clear
distinction between the two models, the main reason being that the Chlorine data 
combine several intermediate and high energy fluxes ($^7$Be, CNO and $^8$B). 
In fact for SSM~I, while the LMA+RSFP 
prediction is fully compatible with the data, the standard LMA one is disfavoured
at 1.3$\sigma$. On the other hand for SSM~II both scenarios 
(LMA and LMA+RSFP) are equally consistent with experimental evidence. 

SNO+ will be able to accurately measure the pep and CNO
fluxes. The former, largely independent of solar models, will supply the
survival probability at low energies, essential to distinguish standard
LMA from LMA+RSFP. Consequently SNO+ will be able to severely constrain the RSFP
interpretation, thus strongly favouring LMA or vice-versa. 
The CNO measurement will on the other hand favour one SSM with respect
to the other. Thus four cases
can be classified according to whether LMA+RSFP or standard LMA is favoured
by the pep measurement and SSM~I or SSM~II is favoured by the CNO one. 
We have seen that, if $\Delta m^2_{01}$ is close to
$1.7\times10^{-8}\ eV$, the value corresponding to the most efficient
resonances, SNO+ will not only be able to discriminate standard LMA from
LMA+RSFP after three years of data taking, but might also discriminate
SSM~II from SSM~I.

Would the RSFP explanation be ruled out - or severely constrained - by the 
pep measurement, and the LMA interpretation favoured, the Chlorine 
results favour the SSM~II solar model, with low heavy element abundance. 
The CNO results could then be used to further confirm the consistency of 
the model.
If, on the other hand, RSFP is favoured by the SNO+ pep-data, CNO can
then be used to identify the right solar model -- indistinguishable 
from Chlorine and high energy neutrino flux data alone.
For the RSFP case considered, if the SSM~I prediction for CNO fluxes 
is found, the SSM~II model can be excluded at  3.5$\sigma$. 
Although it will be harder to discriminate SSM~I from SSM~II, the allowed
regions of SSM~I could be severely constrained.
In any case, other measurements of low energy solar neutrino rates
could complement the identification of the solar model and be used to test
its consistency, and the consistency of the favoured oscillation pattern.

In particular, consistency between the future pep and 
CNO measurements at SNO+ and the Chlorine experiment requires that SNO+ will either
favour an LMA+RSFP scenario or indicate a preference for a low CNO flux prediction 
(as in SSM~II).

\subsection*{Acknowledgements}
The work of MP was partially supported by Funda\c{c}\~{a}o para a
Ci\^{e}ncia e a Tecnologia (FCT) through the grant SFRH/BPD/25019/2005.
The work of NB was supported by FCT through the grant SFRH/BD/28162/2006.
The work of NB and JM was supported by the FCT project grant POCI/FIS/56691/2004.

SA, NB and JM would like to thank M.C.~Chen for many discussions on the potential 
of SNO+ and for access to the results of simulations done at Queen's University.

\end{document}